\documentclass[showpacs,preprintnumbers,onecolumn]{revtex4}

\input{tcilatex}

\begin{document}

\title{Exact treatment of Ising model on the helical tori }
\author{Tsong-Ming Liaw}
\email{ltming@gate.sinica.edu.tw}
\affiliation{Computing Centre, Academia Sinica, 11529 Taipei, Taiwan}
\author{Ming-Chang Huang}
\email{ming@phys.cycu.edu.tw}
\author{Yen-Liang Chou}
\affiliation{Department of physics, Chung-Yuan Christian University, Chungli, Taiwan}
\author{Simon C. Lin }
\affiliation{Institute of Physics, Academia Sinica, 11529 Taipei, Taiwan}
\author{Feng-Yin Li}
\affiliation{Department of Chemistry, National Chung-Hsing University, Taichung, Taiwan }

\begin{abstract}
The exact closed forms of the partition functions of 2D Ising model on
square lattices with twisted boundary conditions are given. The
constructions of helical tori are unambiguously related to the twisted
boundary conditions by virtue of the $SL(2,Z)$ transforms. Numerical
analyses reveal that the finite size effect is irrelevant to the chirality
equipped with each helical boundary condition.
\end{abstract}

\date{\today }
\maketitle


Since Onsager obtained the exact solution of the two-dimensional ($2D$)
Ising model with cylindrical boundary condition ($BC$) in $1944$\cite%
{onsager}, the exact treatments of Ising models on different $2D$ surfaces
have been continuously attempted. Most recently, Wu and Lu \cite{luwu} have
provided analytical treatments for the Ising models with $BC$s of particular
class, including M{\H{o}}bius strip, Klein bottle and self-dual $BC$. The
exact study of the model subject to $BC$s is of fundamental importance.
First, it represents new challenges for the unsolved lattice-statistical
problems \cite{onsager,luwu,kaufman,shmtlb,kw,gh,mcwu,pl,tm,wh}. Second, it
is crucial for the finite-size analysis \cite{fef,mefish,zck,ok,oh}.
Furthermore, it provides an optimal testbed for the predictions of the
conformal field theory \cite{cardyz}. Numerical simulations are plausible
for the exact analyses and provide very rich content for the theory of
finite-size scalings\cite{mefish}. For example, based on the exact analysis
of dimer statistics, by Wu and Lu (\cite{luwu}, $1998$), Kaneda and Okabe 
\cite{ok} have achieved, via computer simulations, more thorough
understanding for the finite-size scaling behaviour of the Ising models
subject to the boundary types of M{\H{o}}bius strip and Klein bottle. While
interesting numerical studies, concerning the excess number of percolation%
\cite{zck} and the Binder parameter\cite{oh}, for the Ising model for the
twisted $BC$s further proceed, the problem for lacking the corresponding
closed form of the partition functions turns out to significant.

Boundary conditions are prescribed by sets of primitive vectors which impose
the periodicity on the corresponding directions. For definite $BC$, sets of
primitive vectors are by no means unique \cite{am}. For $2D$, the equivalent
transformations among the primitive vector-pairs on lattice essentially
preserve the area spanned by the vector-pairs and are thus recognised as $%
SL(2,Z)$. This is the prototype of the modular symmetry discussed in the
context of conformal field theory\cite{cardyz}. The helical $BC$ is of
particular significance owing to its geometrical feature and its relevance
to the formulation for the nanotube physics \cite{sahito}. Helical tori are
formed by pairwise joining the edges of the sheet spanned by any two
orthogonal primitive vectors. The construction ends up with distinct
orientations of the underlying lattice, labelled by the chirality \cite%
{sahito} as well as the chiral aspect ratio. The conventional periodic $BC$
is referred as the helical $BC$ with trivial chirality, as depicted in Fig.
1. The twisted $BC$, on the other hand, counts on the modification to the
conventional $BC$ by cutting the torus and then rejoining after twisting.
Furthermore, the helical $BC$ is shown to be the subclass of twisted one
according to the equivalence relations, as depicted in Fig. 2.

In this Letter, the $2D$ Ising model subject to the twisted $BC$s is exactly
analysed. The general form of such partition functions is obtained, firstly.
Symmetry conditions are employed to reduce the redundancy on setting the
twisting factor $\alpha $ in relation to the conventional aspect ratio $A$.
In addition, any helical $BC$ is shown unambiguously equivalent to a
definite twisted $BC$, by virtue of the $SL(2,Z)$ transform. The finite-size
shift of critical temperature is thus investigated numerically. It turns out
that the scaling behaviour is found chirality-independent. Meanwhile, in
examining the twisting pair dependence, the $A=1$ scaling rule appears to be
twisting-independent. We then conclude by few remarks on the comparison to
the previous numerical issues.

Consider a $M\times N$ square lattice with the coordinates of the lattice
sites specified in form of ${\hat{x}}m+{\hat{y}}n$. The partition function
of the Ising model lattice is given as $Z_{M,N}=[\ 2\cosh (\beta J_{1})\cosh
(\beta J_{2})\ ]^{MN}\ Q_{M,N}$ with the reduced partition function $%
Q_{M,N}=\prod_{m=1}^{M}\prod_{n=1}^{N}\frac{1}{2}{\hat{Q}}_{m,n}$, where ${%
\hat{Q}}_{m,n}=\frac{1}{2}\sum_{\left\{ \sigma _{mn}\right\}
}[(1+t_{1}\sigma _{m,n}\sigma _{m+1,n})(1+t_{2}\sigma _{m,n}\sigma
_{m,n+1})] $. Here we use the notations, $t_{i}=\tanh (\beta J_{i})$ with $%
J_{i}$, for $i=1,2$, denoting the coupling constants along $x$ and $y$
directions, and $\beta =1/k_{B}T$. Twisted $BC$s amount to the
identifications of the spin variables whose locations are related by the
pair of primitive vectors, say, $\overrightarrow{a}_{1}$ and $%
\overrightarrow{a}_{2}$. Basically, only two types of twisting should be in
considerations: One is referred as $Tw_{I}(M,N,d/M)$ specified by the
primitive vectors $\{\overrightarrow{a}_{1}=M{\hat{x}}+d{\hat{y}},%
\overrightarrow{a}_{2}=N{\hat{y}}\}$, and the other is as $Tw_{II}(M,N,d/N)$
specified by $\{\overrightarrow{a}_{1}=M{\hat{x}},\overrightarrow{a}_{2}=d{%
\hat{x}}+N{\hat{y}}\}$. \ 

According to Plechko (\cite{pl}, $1985$), the reduced partition function
takes the form of 
\begin{equation}
Q_{M,N}=\sum_{\left\{ \sigma _{mn}\right\}
}\prod_{m=1}^{M}\prod_{n=1}^{N}\Psi _{m,n}^{(1)}\Psi _{m,n}^{(2)}
\label{eq4}
\end{equation}%
with 
\begin{eqnarray}
\Psi _{m,n}^{(1)} &=&\int da_{m,n}da_{m,n}^{\ast }e^{a_{m,n}a_{m,n}^{\ast
}}[A_{m,n}A_{m+1,n}^{\ast }]  \nonumber \\
\Psi _{m,n}^{(2)} &=&\int db_{m,n}db_{m,n}^{\ast }e^{b_{m,n}b_{m,n}^{\ast
}}[B_{m,n}B_{m,n+1}^{\ast }]  \label{eq5}
\end{eqnarray}%
where $A_{m,n}\ =\ 1+a_{m,n}\sigma _{m,n}$, $A_{m,n}^{\ast }\ =\
1+t_{1}a_{m-1,n}^{\ast }\sigma _{m,n}$, $B_{m,n}\ =\ 1+b_{m,n}\sigma _{m,n}$
and $B_{m,n}^{\ast }\ =\ 1+t_{2}b_{m,n-1}^{\ast }\sigma _{m,n}$, In above,
two pairs of conjugate Grassman variables, $\{a_{m,n},a_{m,n}^{\ast }\}$ and 
$\{b_{m,n},b_{m,n}^{\ast }\}$ have been introduced. 
As technically known to the Refs. \cite{pl,tm,wh}, the handling of the
boundary Boltzmann weights, 
\begin{equation}
\Psi _{\Gamma }=\prod_{n=1}^{N}\Psi _{M,n}^{(1)}\prod_{m=1}^{M}\Psi
_{m,N}^{(2)},  \label{eq7}
\end{equation}%
remains central in the treatments. It turns out to be instructive to
reexamine the paradigm which solves this problem in the original periodic
settings $\sigma _{m+M,n}=\sigma _{m,n}$ and $\sigma _{m,n+N}=\sigma _{m,n}$.

In Ref. \cite{pl} ($1985$), the boundary Boltzmann weights $\Psi _{\Gamma }$
are rearranged such that $\Psi _{\Gamma } =\ \Psi _{\gamma }|_{\Gamma
_{1}}+\Psi _{\gamma }|_{\Gamma _{2}}+\Psi _{\gamma }|_{\Gamma _{3}}-\Psi
_{\gamma }|_{\Gamma _{4}}$ subject to the $BC$s $\Gamma_i$s, imposed on the
Grassman variables, with 
\begin{equation}
\Psi _{\gamma } =\int \prod_{n}^{N}{\overleftarrow{A}_{1,n}^{\ast }}%
\prod_{m}^{M}{\overrightarrow{B}_{m,1}^{\ast }}\prod_{n}^{N}{\overrightarrow{%
A}_{M,n}}\prod_{m}^{M}{\overleftarrow{B}_{m,N}},  \label{eq7_1}
\end{equation}%
where the arrows indicate the ordering for the multiplications and we employ
the notation $\int $ for all the coming weighted integration over relevant
Grassman variables. Subsequently, mirror ordering is applied routinely and
furnishes the simple expression of pure Grassmanian integrations, 
\begin{eqnarray}
Q_{M,N} &=&\frac{1}{2}[G|_{\Gamma _{1}}+G|_{\Gamma _{2}}+G|_{\Gamma
_{3}}-G|_{\Gamma _{4}}],  \label{eq9} \\
G &=&\int \exp \{\sum_{m,n}^{M,N}[a_{m,n}b_{m,n}+t_{1}t_{2}a_{m-1,n}^{\ast
}b_{m,n-1}^{\ast }+(t_{1}a_{m-1,n}^{\ast }+t_{2}b_{m,n-1}^{\ast
})(a_{m,n}+b_{m,n})]\},  \label{eq10}
\end{eqnarray}%
where the integrations can be diagonalised and carried out, straightforward 
\cite{pl}.

The reviewing paragraph above suggests that the modification is only
essential for the twisted $BC$ in the the key steps, i.e., from Eq. (\ref%
{eq7}) to Eq. (\ref{eq7_1}). For $Tw_{I}$, the $\Psi _{M,n}^{(1)}$ term in
Eq. (\ref{eq7}) is calibrated in relation to the toroidal one. This then
leads to 
\begin{equation}
\Psi _{\Gamma }=\int {\prod_{m=1}^{M}\overrightarrow{B_{m,1}^{\ast }}%
\prod_{k=1}^{N-d}\overleftarrow{(1-t_{1}a_{M,k+d}^{\ast }\sigma _{1,k})}%
\prod_{k=n-d+1}^{N}\overleftarrow{(1-t_{1}a_{M,k+d-N}^{\ast }\sigma _{1,k})}}%
\prod_{n=1}^{N}\overrightarrow{A_{M,n}}\prod_{m=1}^{M}\overleftarrow{B_{m,N}}%
,  \label{eq11_0}
\end{equation}%
where the $BC$ $\sigma _{m+M,n+d}=\sigma _{m,n}$ has been explicitly
employed. Reordering of the first three products in Eq. (\ref{eq11_0}) is
essential such that the form of Eq. (\ref{eq7_1}) can be achieved. By
recursive use of the identity for the permutation of Grassmanian functions%
\cite{pl}, we employ, instead, 
\begin{equation}
XYZ\equiv \frac{1}{2}(ZYX^{-}-Z^{-}Y^{-}X^{-}+ZY^{-}X+ZY^{-}X^{-}),
\label{eq12}
\end{equation}
where $X, Y$ and $Z$ stand for the corresponding three objects and the
superscript "$-$" denotes flipping the sign of the Grassman variables.
Accordingly, the form of Eq. (\ref{eq7_1}) is achieved which implies Eq. (%
\ref{eq9}). Note that the form of Eq. (\ref{eq10}) preserves under twisting.
However, the $BC$s imposed on the Grassman variables are modified in
response to the corresponding sign flipping appearing in the deduction of
Eq.(\ref{eq12}). For convenience, the compact notation as $\Gamma
_{i}=\left( \pm ,\pm \right) $ can be employed as follows. The first sign in
the parenthesis corresponds to $a_{m,N}^{\ast }=\pm \ a_{m,0}^{\ast }$ and
the second one is for $a_{M,n+d}^{\ast }=\pm \ a_{0,n}^{\ast }$. The $BC$s
are given as $\Gamma _{1}=\left( -,-\right) $, $\Gamma _{2}=\left(
+,-\right) $, $\Gamma _{3}=\left( -,+\right) $ and $\Gamma _{4}=\left(
+,+\right) $. The exact partition function is straightforward, henceforth.

For $Tw_I$, with $\alpha\equiv d/M$, the reduced partition function is 
\begin{eqnarray}
Q_{M,N}^{\alpha }&=&\frac{1}{2}\left\{ I_{M,N}^{\alpha }(\frac{1}{2},\frac{1%
}{2})+I_{M,N}^{\alpha }(\frac{1}{2},0)+I_{M,N}^{\alpha }(0,\frac{1}{2})-%
\mbox{sgn}(\frac{T-T_{c}}{T_{c}})I_{M,N}^{\alpha }(0,0)\right\},
\label{eq13} \\
I_{M,N}^{\alpha }(\Delta ,{\bar{\Delta}})&=&\prod_{p=1}^{M}\prod_{q=1}^{N}%
\left\{ \lambda _{0}-\lambda _{1}\cos \left[ 2\pi \left( \frac{p+\Delta }{M}-%
\frac{\alpha \left( q+{\bar{\Delta}}\right) }{N}\right) \right] -\lambda
_{2}\cos \left[ 2\pi \left( \frac{q+{\bar{\Delta}}}{N}\right) \right]
\right\} ^{1/2},  \label{eq14}
\end{eqnarray}%
where $\lambda _{0}=(1+t_{1}^{2})(1+t_{2}^{2})$,$\ \lambda
_{1}=2t_{1}(1-t_{2}^{2})$ and $\lambda _{2}\ =\ 2t_{2}(1-t_{1}^{2})$. In
addition, the function $\mbox{sgn}(x)$ denotes the sign of the value $x$ and 
$T_{c}$ is the critical temperature of the bulk system. The reduced
partition function for $Tw_{II}$ remains formally as Eq. (\ref{eq13} ) but
with 
\begin{equation}
I_{M,N}^{\alpha }(\Delta ,{\bar{\Delta}})=\prod_{p=1}^{M}\prod_{q=1}^{N}%
\left\{ \lambda _{0}-\lambda _{1}\cos \left[ 2\pi \left( \frac{p+\Delta }{M}%
\right) \right] -\lambda _{2}\cos \left[ 2\pi \left( \frac{q+{\bar{\Delta}}}{%
N}-\frac{\alpha \left( p+\Delta \right) }{M}\right) \right] \right\} ^{1/2},
\label{eq15}
\end{equation}%
where, instead, $\alpha=d/N$.

It can be checked for Eq. (\ref{eq13}) that $Q_{M,N}^{\alpha
}=Q_{M,N}^{-\alpha }$ based on either Eq. (\ref{eq14}) or Eq. (\ref{eq15}),
while, intuitively, twisting either clockwise or counterclockwise is not
classified by the system. Noteworthy is also that reversing the sign of a
twist factor $\alpha $ can not be obtained via the $SL(2,Z)$ transform. On
employing this transform explicitly, pairs of primitive vectors are related
among each other in the manner of 
\begin{equation}
\left( 
\begin{array}{c}
{\overrightarrow{a}}_{1}^{^{\prime }} \\ 
{\overrightarrow{a}}_{2}^{^{\prime }}%
\end{array}%
\right) =\mathcal{M}\left( 
\begin{array}{c}
{\overrightarrow{a}}_{1} \\ 
{\overrightarrow{a}}_{2}%
\end{array}%
\right) \ \forall \ \mathcal{M}\ \in SL(2,Z).  \label{eq16}
\end{equation}%
Consider $Tw_I$, for example. The choice of matrix elements $\mathcal{M}%
_{11}=1$, $\mathcal{M}_{12}=J\in Z$, $\mathcal{M}_{21}=0$ and $\mathcal{M}%
_{22}=1$ gives rise to the new pairs of primitive vectors $\{\overrightarrow{%
a}_1=M{\hat x}+(d+N){\hat y}, \overrightarrow{a}_2=N{\hat y} \}$, which
prescribes the same $BC$. As evidence, $Q_{M,N}^{\alpha }=Q_{M,N}^{\alpha
+JA}$ can be explicitly checked, where the conventional aspect ratio appears
as $A=N/M$. Therefore, the effective range of $\alpha $ is $0\leq \alpha <A$%
. In addition, the equivalence $Tw_{I}( M,N,\alpha =A/J) \cong \ Tw_{II}(
M^{^{\prime }}=JM,N^{^{\prime }}=N/J,\alpha ^{^{\prime }}=1/\alpha ) $ can
be achieved by virtue of choosing the elements $\mathcal{M}_{11}=J\in Z$, $%
\mathcal{M}_{12}=-1$, $\mathcal{M}_{21}=1$ and $\mathcal{M}_{22}=0$ with $%
\alpha =A/J$ and $\alpha ^{^{\prime }}=J/A$. 
Again, the partition functions based on Eqs. (\ref{eq14} and (\ref{eq15})
appear to fulfil the these relations. Hence, it is sufficient to study the
unique correspondence of a helical torus to the one of the above twistings,
say $Tw_{I}$.

The helical tori, on the hand, lie in the orthogonal primitive vector pair, 
\begin{eqnarray}
{\overrightarrow{a}}_{1}^{\prime } &=&{\hat{x}}\ P_{1}+{\hat{y}}\ Q_{1}, 
\nonumber \\
{\overrightarrow{a}}_{2}^{\prime } &=&-{\hat{x}}\ Q_{2}+{\hat{y}}\ P_{2},
\label{eq17}
\end{eqnarray}%
where the two radii for the torus are given as $L_{i}=\sqrt{%
P_{i}^{2}+Q_{i}^{2}}$ for $i=1,2$. Hence, let the helical system denoted by $%
Hl(B,L_{1},\chi )$, where the chiral aspect ratio $B=L_{2}/L_{1}$ and the
chirality $\chi =Q_{1}/P_{1}\equiv \ Q_{2}/P_{2}$. In order to furnish the
equivalent structure $Hl(B,L_{1},\chi )\ \cong \ Tw_{I}(A,M,\alpha )$, $%
\mathcal{M}_{11}={P_{1}}/{M}$ and $\mathcal{M}_{21}=-{Q_{2}}/M$ implies that 
\begin{eqnarray}
\mathcal{M}_{21} &=&-B\chi \mathcal{M}_{11}  \label{eq17_1} \\
1 &=&\mathcal{M}_{11}\mathcal{M}_{22}\ -\ \mathcal{M}_{21}\mathcal{M}_{12}.
\label{eq18} \\
A &=&\frac{\left( \mathcal{M}_{21}\right) ^{2}}{B}+B\left( \mathcal{M}%
_{11}\right) ^{2},  \label{eq19} \\
\alpha &=&-\frac{\mathcal{M}_{21}\mathcal{M}_{22}}{B}-B\mathcal{M}_{11}%
\mathcal{M}_{12}.  \label{eq20}
\end{eqnarray}%
However, few remarks on the uniqueness of the relations above remain
essential.

The $[\mathcal{M}_{12},\mathcal{M}_{22}]$ pair is unambiguously determined
up to $\mathcal{M}_{11}$ and $\mathcal{M}_{21}$ for $0\leq \alpha <A$. This
is because shifting $[\mathcal{M}_{12},\mathcal{M}_{22}]$ by appending $[J%
\mathcal{M}_{11},J\mathcal{M}_{21}]$ leaves Eq.(\ref{eq18}) invariant $%
\forall J\in Z$but only causes $\alpha $ deviated by $JA$ in Eq. (\ref{eq20}%
). Meanwhile, the allowable region for $[\mathcal{M}_{12}$, $\mathcal{M}%
_{22}]$ appropriate for $0\leq \alpha <A$ is exactly one vector section $[%
\mathcal{M}_{11},\mathcal{M}_{21}]$. In addition, the ambiguity relating to
size dependence can be removed by noting the coprime properties between $%
\mathcal{M}_{11}$ and $\mathcal{M}_{21}$ which ensures the solubility of the
integer pair $[\mathcal{M}_{12},\mathcal{M}_{22}]$ subject to Eq. (\ref{eq18}%
). Consequently, a helical two-tuple $(B,\chi )$ is equipped a unique pair $%
\{A,\alpha \}$ for the twisting in the effective range. Moreover, the
classification of helical via the twisting parameters remains unambiguous.
This is because if different helical tori were equivalent to the same
twisted $BC$, SL(2,Z) transforms would have been held among them, which can
be shown impossible. The effective range of the helical $BC$ can also be
further reduced. The partition function is unable to classify the rolling up
direction in forming the tori, hence, no distinction between the
characterisations $\chi $ and $-\chi$ is essential. Subsequently, once $%
Hl(B,\chi )\ \cong \ Tw_{I}(A,\alpha )$, one can derive that $Hl(1/B,1/\chi
)\cong \ Tw_{I}(A,-\alpha )$. To be concrete, assuming $\chi>0$, $%
\chi^{\prime}=1/\chi$ implies $\mathcal{M}_{11}^{\prime }=-\mathcal{M}_{21}, 
\mathcal{M}_{21}^{\prime }=-\mathcal{M}_{11}$ as well as $\mathcal{M}%
_{12}^{\prime }=\mathcal{M}_{22}$ and $\mathcal{M}_{22}^{\prime }=\mathcal{M}%
_{12}$ according to Eq. (\ref{eq17_1}). This then preserves $A$ but gives $%
\alpha^{\prime }=-\alpha $ by virtue of Eqs. (\ref{eq18}) and (\ref{eq19}).

The shift of the specific-heat peak $T_{\max }$ away from the critical
temperature $T_{c}$ under the isotropic couplings can be computed from the
the exact partition function. Upon using the parametrisation in terms of $%
Tw_{I}(A,M,\alpha )$, the critical shift $\theta (A,\alpha )=(T_{\max
}(\alpha )-T_{c})/T_{c}$ is plotted against $1/L=1/\sqrt{MN}$ in Fig. 3. For
helical $BC$s, Eqs. (\ref{eq17_1})-(\ref{eq20}) are employed in order to
determine the critical shifts $\theta (B,\chi )$ versus $1/L=1/\sqrt{%
L_{1}L_{2}}=1/\sqrt{MN}$ for various $B$s and $\chi $s, as shown in Fig. 4.
In Fig.3, the curves of scaling for definite $A$ deviate by altering the
twist factor $\alpha $. However, no such splitting is found for the
exceptional case $A=1$. On the other hand, all the critical shifts $\theta
\left( B,\chi \right) $ with the same $B$ value, in Fig. 4, fall into to one
single smooth curve, the finite-size effect turns out to be
chirality-independent. Hence, the matching of the two particular curves,
i.e. for $B=1$ in Fig.3 and for $A=1$ in Fig. 4, appears to be an additional
feature. Moreover, the critical shift $\theta \left( B,\chi \right) $ flips
its sign at $B=b_{0}$ and $1/b_{0}$ with $b_{0}\simeq 3$, as it was
anticipated by Ferdinand and Fisher \cite{fef} for the conventional periodic
BC, where the exact $b_{0}$ value was determined as $b_{0}=3.13927..$, a
result which now applies for all the helical tori.

In conclusion, we provide the complete description for the finite-size
effect of Ising Model subject to the subclass helical $BC$s of the twisted
tori. This is explicitly done by solving the exact form of the partition
function appropriate for all the twisted $BC$s. The evidence of the
finite-size effect being chirality-independent basically supports the
invariance of the scaling behaviour of partition function under rotation of
the primitive vector pair subject to $BC$s, conjectured in Ref. \cite{oh}.
However, the particular coincidence for $A=1$ and $B=1$ regardless $\alpha$
and $\chi$ suggests further interesting points exceeding beyond the
rotational invariance. For consistency, we stress the fact that $A=1$ does
not non-trivially permits any helical structure, as one may observe in Eqs. (%
\ref{eq17_1})-(\ref{eq20}). As the final remark, the invariant aspect ratio%
\cite{oh} $A/(1+\alpha ^{2})$ \cite{fono} coincides with $B$ only for $\chi
=\alpha $, nor does it pertain to the case where $A=1$.

This work was partially supported by the National Science Council of
Republic of China (Taiwan) under the Grant No. NSC 93-2212-M-033-005.

FIG. 1. The formation of helical tori by pairwise joining the edges of the
rectangle spanned by any orthogonal set of vectors on the lattice plane: (a)
the direction of the primitive vectors coincides with the lattice
orientations for the conventional toroidal $BC$ and (b) the helical tori are
formed for the non-coincidence.

FIG. 2. Equivalence between the $BC$s in helical and twisted schemes
prescribed by $\{{\vec{a}}_{1},{\vec{a}}_{2}\}$ and $\{{\vec{a}}_{1}^{\prime
},{\vec{a}}_{2}^{\prime }\}$ respectively, on a $M\times N$ square lattice.
For the helical $BC$, the setting $Q_{1}/P_{1}=Q_{2}/P_{2}$ ensures that the
two primitive vectors are orthogonal. On the other hand, twisting is
generated by a $d$-unit traverse shift.

FIG. 3. Plotting $\theta (A,\alpha )$ against $1/L$ for $A=1,2,3,4$ with $%
\alpha =0({\diamond }),0.1A(\bigtriangleup ),0.2A(\bigtriangledown
),0.3A(\circ ),0.4A({\bullet })$ and $0.5A(\times )$. The scaling behaviours
are obviously deviated by $\alpha $. Nonetheless, for $A=1$ no splitting is
found with respect to the twisting factors.

FIG. 4. The plot of $\theta (B,\chi )$ versus $1/L$. For a given chiral
aspect ratio $B$. Results of different chiralities $\chi $ collapse into one
curve and the curves of both $\theta (B,\chi )$ and $\theta (1/B,1/\chi )$
versus $1/L$ coincide.

\end{document}